# A Landau-Devonshire Analysis of Strain Effects on Ferroelectric Al$_{1-x}$Sc$_x$N

*Keisuke Yazawa$^{1,2*}$, Andriy Zakutayev$^1$, and Geoff L. Brennecka$^{2*}$*

1. Materials Science Center, National Renewable Energy Laboratory, Golden, Colorado 80401, United States

2. Department of Metallurgical and Materials Engineering, Colorado School of Mines, Golden, Colorado 80401, United States

**Corresponding Authors**

*E-mail: Keisuke.Yazawa@nrel.gov and geoff.brennecka@mines.edu

**ABSTRACT**

We present a thermodynamic analysis of the recently discovered nitride ferroelectric materials using the classic Landau-Devonshire approach. The electrostrictive and dielectric stiffness coefficients of Al$_{1-x}$Sc$_x$N with wurtzite structure (*6mm*) are determined using a free energy density function assuming a hexagonal parent phase (*6/mmm*), with the first order phase transition based on the dielectric stiffness relationships. The results of this analysis show that the strain sensitivity of the energy barrier is one order of magnitude larger than that of the spontaneous polarization in these novel wurtzite ferroelectrics, yet both are less sensitive to strain compared to classic perovskite ferroelectrics. These analysis results reported here explain experimentally reported sensitivity of coercive field to elastic strain/stress in Al$_{1-x}$Sc$_x$N films, and would enable further thermodynamic analysis via phase field simulation and related methods.

## I. INTRODUCTION

Landau – Devonshire thermodynamic modelling has been widely acknowledged as a valuable tool for the phenomenological description of the physics of ferroelectrics.[1] The approach has been validated across a wide range of ferroelectric materials from the first ferroelectric discovered, Rochelle Salt, to recent HfO$_2$-based materials.[2–4] Extension to what is now referred to as the Landau – Ginsburg – Devonshire model enables rigorous investigation of polarization boundaries and reorientation, phase transformations, and domain formation/evolution.[5–10] Most recently, full ranges of the double well function, an important outcome of the Landau – Devonshire theory, have been experimentally observed including the unstable "spinodal" region[11] associated with negative capacitance.[12,13] Thus, the phenomenological Landau – Devonshire theory continues to be a valuable tool for understanding and predicting properties and behaviors of ferroelectric materials.

Since the discovery of Al$_{1-x}$Sc$_x$N ferroelectrics,[14] investigation of wurtzite ferroelectrics has been of great interest due to their large spontaneous polarization values and process compatibility with existing Si technology.[15–22] One challenge with these materials is the large coercive field. In this context, thermodynamic analysis is an attractive approach to predict and understand changes in properties under thermodynamic variables such as strain, electric field, and temperature. However, most of the thermodynamic coefficients have not yet been determined for the wurtzite ferroelectrics. Most recently, Wang, et al. described the thermodynamic double well function for Al$_{1-x}$Sc$_x$N using DFT calculation results for the second and fourth order dielectric stiffness.[23] Although that effort is an important start, higher order terms are critical for some materials,[24,25] and coupling terms that are non-negligible in many ferroelectric systems have not yet been addressed. Indeed, limited ferroelectric property tuning under elastic strain/stress has been reported in Al$_{1-x}$Sc$_x$N,[14,16,26] but consistent descriptions remain elusive.

The Landau-Devonshire free energy density (hereafter, "density" will be omitted for brevity) of a ferroelectric material with a centrosymmetric parent phase, whose variables are polarization $P_i$ and total strain $\varepsilon_{ij}$ can be expressed as[27]

$$f = a_{ij}P_iP_j + a_{ijkl}P_iP_jP_kP_l + a_{ijklmn}P_iP_jP_kP_lP_mP_n + \frac{1}{2}c_{ijkl}\varepsilon_{ij}\varepsilon_{kl} - q_{ijkl}\varepsilon_{ij}P_kP_l \quad (1)$$

where $a_{ij}$, $a_{ijkl}$ and $a_{ijklmn}$ are the dielectric stiffness coefficients at constant strain, $c_{ijkl}$ is the elastic stiffness at constant polarization, and $q_{ijkl}$ is the electrostrictive coefficient. Note that the summation symbols for each term are omitted hereafter. In a stress-free state, the strain derivative of equation (1) is 0, namely[7,28,29]

$$\frac{\partial f}{\partial \varepsilon_{ij}} = \sigma_{ij} = c_{ijkl}\varepsilon_{kl}^0 - q_{ijkl}P_kP_l = 0 \quad (2)$$

and the strain satisfying the relationship is the spontaneous strain $\varepsilon_{kl}^0$, which is expressed as

$$\varepsilon_{kl}^0 = s_{klmn}q_{mnop}P_oP_p = Q_{klop}P_oP_p \quad (3)$$

where $Q_{ijkl}$ is polarization – strain electrostrictive coefficient. Plugging equation (3) into equation (1) gives the free energy at the stress-free boundary condition $f^{sf}$, which is expressed as[7,28,29]

$$f^{sf} = a_{ij}P_iP_j + a^*_{ijkl}P_iP_jP_kP_l + a_{ijklmn}P_iP_jP_kP_lP_mP_n \quad (4)$$

where $a^*_{ijkl}$ is expressed as

$$a^*_{ijkl} = a_{ijkl} - \frac{1}{2}c_{mnop}Q_{mnij}Q_{opkl} \quad (5)$$

From this representation of the free energy under stress-free conditions, the polarization reorientation in a relaxed crystal at any point during the switching event can be discussed.

In this work, we apply this classic approach to the recently discovered wurtzite nitride ferroelectric $Al_{1-x}Sc_xN$ and are able to explain the origins of experimentally-observed changes in coercive field with elastic strain while also predicting a first order phase transition to the centrosymmetric prototype phase. We use Landau – Devonshire thermodynamic modeling to describe elastic strain effects on $Al_{1-x}Sc_xN$ ferroelectrics up to $6^{th}$ order polarization terms, including elastic and electromechanical coupling terms. The coefficients are determined based upon a range of DFT calculations and experimental results reporting electric, elastic, and piezoelectric properties. The electrostrictive coefficient and range of dielectric stiffness values are determined for $Al_{1-x}Sc_xN$. We find that—consistent with experimental reports—the coercive field is more sensitive than spontaneous polarization to elastic strain; we also determine that strain has a significantly smaller effect on all relevant properties of $Al_{1-x}Sc_xN$ than the classic perovskite $PbTiO_3$.

## II. MODEL PARAMETERS FOR $Al_{1-x}Sc_xN$

### A. Dielectric Stiffness

In wurtzite ferroelectrics, the *6/mmm* hBN structure is assumed to be the reference parent structure.[14,30,31] Based on equation (4), the stress-free free energy $f^{sf}$ can be explicitly written to account for coefficient degeneracy due to the crystal symmetry as:

$$f^{sf} = a_{11}(P_1^2 + P_2^2) + a_{33}P_3^2 + a_{1111}^*(P_1^4 + P_2^4) + a_{3333}^*P_3^4 + a_{1122}^*P_1^2P_2^2$$
$$+ a_{1133}^*(P_1^2P_3^2 + P_2^2P_3^2) + a_{111111}(P_1^6 + P_2^6) + a_{333333}P_3^6$$
$$+ a_{111122}(P_1^4P_2^2 + P_1^2P_2^4) + a_{111133}(P_1^4P_3^2 + P_1^2P_3^4 + P_2^4P_3^2 + P_2^2P_3^4)$$
$$+ a_{112233}P_1^2P_2^2P_3^2 \tag{6}$$

The spontaneous polarization in wurtzite is along the 3 direction. Under conditions with no in-plane polarization in the system ($P_1 = P_2 = 0$, $P_3 \neq 0$), namely, when there exists no in-plane electric field or shear strain, conditions compatible with a polar oriented fiber-textured film or single crystal, the stress-free free energy can be reduced to

$$f^{sf} = a_{33}P_3^2 + a_{3333}^*P_3^4 + a_{333333}P_3^6 \tag{7}$$

The local minimum of the free energy represents the spontaneous polarization $P_s$, thus

$$\left.\frac{\partial f^{sf}}{\partial P_3}\right|_{P_3=P_s} = 0 \tag{8}$$

Also, the second derivative of the free energy, which corresponds to the inverse of the slope of the *P-E* curve, is the observed dielectric susceptibility along the wurtzite 3 direction, namely

$$\left.\frac{\partial^2 f^{sf}}{\partial P_3^2}\right|_{P_3=P_s} = \frac{1}{\kappa_0\chi_{33}} \tag{9}$$

where $\kappa_0$ and $\chi_{33}$ are the vacuum permittivity and material susceptibility, respectively. From differential equations (8) and (9), the higher order dielectric stiffness coefficients $a_{3333}^*$ and $a_{333333}$ are expressed in terms of $a_{33}$ as,

$$a_{3333}^* = -\frac{8a_{33}\kappa_0\chi_{33} + 1}{8P_s^2\kappa_0\chi_{33}} \tag{10}$$

$$a_{333333} = -\frac{4a_{33}\kappa_0\chi_{33} + 1}{12P_s^4\kappa_0\chi_{33}} \tag{11}$$

Thus, by plugging equations (10) and (11) into (7), the polarization – free energy curve at the stress free condition can be illustrated with only one unknown variable, $a_{33}$, and experimentally or computationally obtained dielectric constant and spontaneous polarization. Fig. 1 shows the polarization – free energy for stress-free conditions across a wide range of various $a_{33}$ values for Sc content $x = 0.2$ in $Al_{1-x}Sc_xN$. The range is limited to the negative value of $a_{33}$ to focus on the ferroelectric region.[27] The spontaneous polarization and susceptibility values utilized here are from reported DFT calculation results,[32,33] though they are consistent with a number of experimental reports as well.[14,15,33] The second order term starts significantly affecting the curve shape when $a_{33} < -1 \times 10^9$ J m C$^{-2}$, and the well depth gets deeper with decreasing $a_{33}$ value. This well depth relates to the intrinsic ferroelectric coercive field, which is difficult to determine unambiguously due to kinetic contributions (e.g., frequency dependence of coercive field[34]) and strong influence of defects and extrinsic contributions to polarization reversal.[35,36] However, the lower limit of the $a_{33}$ value can be determined via the curve shape shown in Fig. 1. In one limiting case for $a_{33} = -1 \times 10^{10}$ J m C$^{-2}$, the free energy decreases again for $P > 1.6$ C m$^{-2}$ due to the strong negative contribution of the second order term to the free energy. This is irrational because infinite polarization becomes the most stable state, which physically corresponds to decomposition via cation/anion displacement. Thus, stability dictates that the limit of $f^{sf}$ of $P$ as $P$ approaches $\infty$ needs to be positive, namely

$$\lim_{P \to \infty} f^{sf} = \frac{\infty(4a_{33}\kappa_0\chi_{33} + 1)}{\text{sgn}(\kappa_0\chi_{33})} > 0 \tag{12}$$

From this, the $a_{33}$ range is simply determined as

$$a_{33} > -\frac{1}{4\kappa_0\chi_{33}} \tag{13}$$

Also, the sign of $a^*_{3333}$ determines the order of the ferroelectric – dielectric phase transition.[27] From the $a_{33}$ - $a^*_{3333}$ relationship written in equation (10), the phase transition order is determined as follows,

$$\begin{cases} a_{33} > -\dfrac{1}{8\kappa_0 \chi_{33}} & \text{... First order phase transition} \\ a_{33} < -\dfrac{1}{8\kappa_0 \chi_{33}} & \text{... Second order phase transition} \end{cases} \quad (14)$$

Fig. 1(b) illustrates the $a_{33}$ lower limit and the first and second order transition range for the Sc range from $x = 0 - 0.5$. From a DFT calculated barrier height, the $a_{33}$ value and phase transition order can be determined. Wang, et al. reported 0.12 eV/f.u. and 0.23 eV/f.u. for the barrier height of the double well function at $x = 0.2$ and 0 in a lattice relaxed transition that corresponds to the stress-free transition in this discussion.[23] Based on the DFT calculated barrier height, $a_{33}$ at $x = 0.2$ is -0.32 GJ m C$^{-2}$, which is in the first order phase transition range shown in Fig. 1(b). It is noteworthy that the $a_{33}$ value is comparable to that of classic perovskite ferroelectrics, such as -0.027 GJ m C$^{-2}$ for BaTiO$_3$ and -0.17 GJ m C$^{-2}$ for PbTiO$_3$.[29,37]

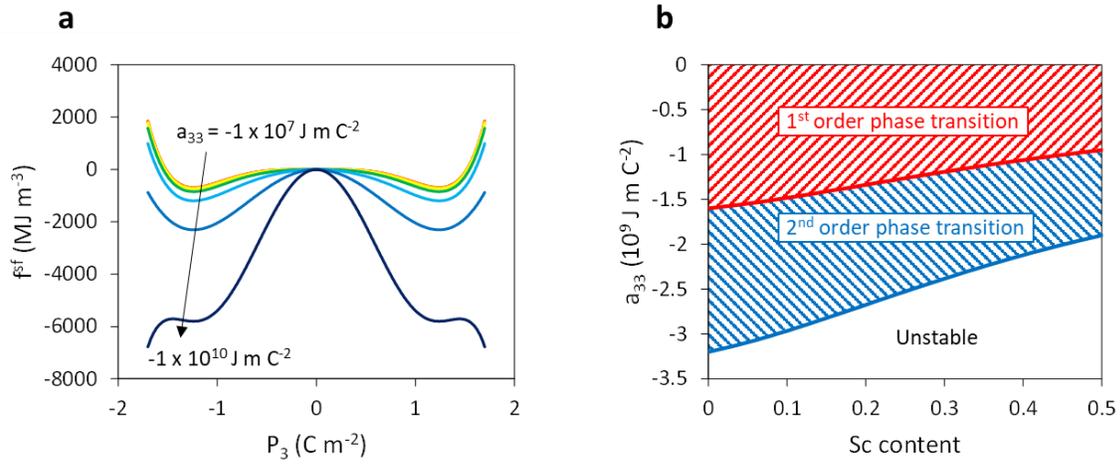

FIG. 1 Second order dielectric stiffness $a_{33}$ analysis (a) Free energy curves with various $a_{33}$ values for Sc content = 0.2. (b) First and second order phase transition range in $a_{33}$ as a function of Sc content.

## B. Electrostrictive Coefficient

To understand the electro-mechanical coupling contribution to the spontaneous polarization, the electrostrictive coefficient needs to be determined. In the *6mm* wurtzite material system, the electrostrictive and piezoelectric coefficients are related as follows,[1,38]

$$d_{311} = d_{322} = 2Q_{1133}P_s\chi_{33} \quad (15)$$

$$d_{333} = 2Q_{3333}P_s\chi_{33} \quad (16)$$

Besides the reported spontaneous polarization and susceptibility,[32,33] DFT simulation results for the piezoelectric coefficient[39,40] are used to calculate the electrostrictive coefficient. Fig. 2 shows the electrostrictive coefficient as a function of Sc content. The absolute values of both $Q_{3333}$ and $Q_{1133} = Q_{2233}$ increase with increasing Sc content, similar to the $Q$ and $d$ increase in classic ferroelectric solid solutions such as $PbZrO_3$ – $PbTiO_3$.[33,39,41] It is noteworthy that the $Q$ values at Sc content $x = 0.5$ (e.g., $Q_{3333} = 0.247$ m$^4$ C$^{-2}$) are significantly higher than those at the MPB of $Pb(Zr,Ti)O_3$ (e.g., $Q_{3333} = Q_{1111} = 0.097$ m$^4$ C$^{-2}$), which corresponds to the abrupt increase of piezoelectric coefficient in larger Sc content.[41,42] Polynomial fits from the composition – electrostrictive relationships are

$$Q_{3333} = 4.1068x^3 - 1.7514x^2 + 0.2846x + 0.0226 \quad (17)$$

$$Q_{1133} = 1.3904x^3 + 0.6022x^2 - 0.1021x - 0.0105 \quad (18)$$

where *x* is the Sc content ranging from 0 to 0.5. The dielectric stiffness and electrostrictive coefficient values for $x = 0$ to 0.5 are tabulated in Table 1.

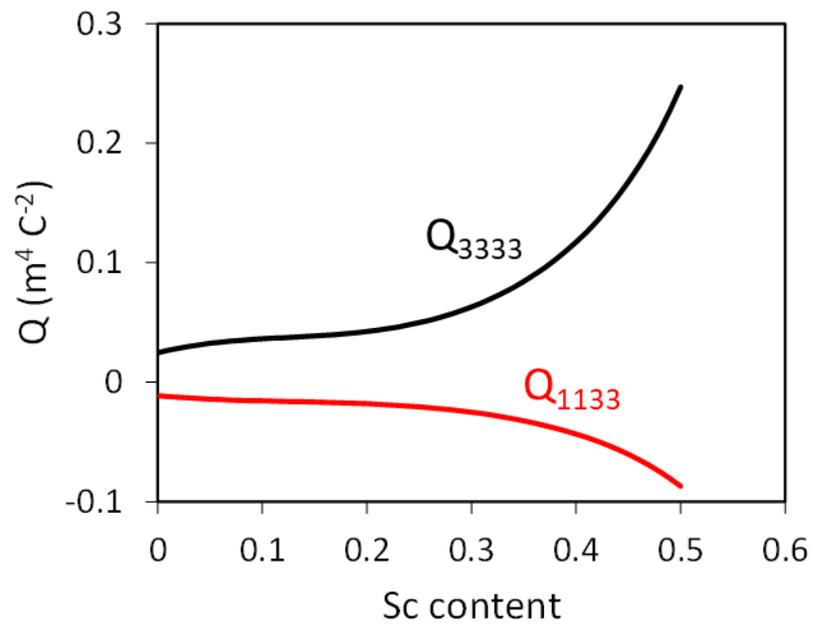

FIG. 2 Electrostrictive coefficient as a function of Sc content.

Table I. Range of dielectric stiffnesses and electrostrictive coefficients for x = 0 to 0.5

| | Equation | Unit | Sc content $x$ in Al$_{1-x}$Sc$_x$N | | | | | |
| --- | --- | --- | --- | --- | --- | --- | --- | --- |
| | | | 0* | 0.1 | 0.2* | 0.3 | 0.4 | 0.5 |
| $a_{33}$ | $> -\dfrac{1}{4\kappa_0 \chi_{33}}$ | GJ m C$^{-2}$ | -1.84 | >-2.97 | -0.32 | >-2.38 | >-2.12 | >-1.90 |
| $a^*_{3333}$ | $= -\dfrac{8a_{33}\kappa_0\chi_{33}+1}{8P_s^2\kappa_0\chi_{33}}$ | GJ m$^5$ C$^{-4}$ | 0.33 | <0.92 | -0.66 | <0.84 | <0.89 | <1.23 |
| $a_{333333}$ | $= -\dfrac{4a_{33}\kappa_0\chi_{33}+1}{12P_s^4\kappa_0\chi_{33}}$ | GJ m$^9$ C$^{-6}$ | 0.12 | >0 | 0.33 | >0 | >0 | >0 |
| $a_{3333}$ | $= a^*_{3333} + \tfrac{1}{2}(2c_{1111}Q^2_{1133} + 2c_{1122}Q^2_{1133} + 4c_{1133}Q_{1133}Q_{3333} + c_{3333}Q^2_{3333})$ | GJ m$^5$ C$^{-4}$ | 0.55 | <1.13 | -0.44 | <1.19 | <1.70 | <3.28 |
| $Q_{3333}$ | -- | m$^4$ C$^{-2}$ | 0.025 | 0.036 | 0.043 | 0.063 | 0.117 | 0.247 |
| $Q_{1133}$ | -- | m$^4$ C$^{-2}$ | -0.011 | -0.016 | -0.018 | -0.025 | -0.043 | -0.087 |

*The dielectric stiffnesses are determined by energy barrier height of a DFT result at $x = 0$ and 0.2[26]

## IV. ELECTROMECHANICAL COUPLING EFFCTS

### A. Formulation

With the determined electrostrictive coefficient, elastic strain - spontaneous polarization coupling can be investigated. The explicit expression of equation (1) including the total strain as a variable can be simplified considering the crystal symmetry and polarization direction to,

$$f = a_{33}P_3^2 + a_{3333}P_3^4 + a_{333333}P_3^6 + \frac{1}{2}c_{1111}(\varepsilon_{11}^2 + \varepsilon_{22}^2) + \frac{1}{2}c_{3333}\varepsilon_{33}^2 + c_{1122}\varepsilon_{11}\varepsilon_{22}$$
$$+ c_{1133}(\varepsilon_{11}\varepsilon_{33} + \varepsilon_{22}\varepsilon_{33}) - q_{1133}(\varepsilon_{11} + \varepsilon_{22})P_3^2 - q_{3333}\varepsilon_{33}P_3^2$$

Note that shear strain contributions to the free energy are omitted here due to the assumption that no in-plane polarization emerges. The 4th order dielectric stiffness, $a_{3333}$, under fixed strain boundary conditions, can be back-engineered from equation (5) (with symmetry considerations) as,

$$a_{3333} = a_{3333}^* + \frac{1}{2}(2c_{1111}Q_{1133}^2 + 2c_{1122}Q_{1133}^2 + 4c_{1133}Q_{1133}Q_{3333} + c_{3333}Q_{3333}^2) \qquad (20)$$

Also, $q_{1133}$ and $q_{3333}$ can be represented in terms of $Q_{1133}$ and $Q_{3333}$, and reported stiffness $c_{ijkl}$ even though the other $Q_{ijkl}$ components are unknown,[39] namely

$$q_{1133} = c_{11mn}Q_{mn33} \qquad (21)$$
$$q_{3333} = c_{33mn}Q_{mn33} \qquad (22)$$

Although the entire free energy curve as a function of polarization at a fixed strain state represents a rare case (e.g., hydrostatically-constrained crystal), the local minima give the spontaneous polarization at a static state with an arbitrary strain. Thus, spontaneous polarization $P^*_s(\varepsilon)$ satisfies the derivative equation

$$\left.\frac{\partial f}{\partial P_3}\right|_{P_3=P_s^*} = 0 \qquad (23)$$

namely, the spontaneous polarization as a function of total strain is expressed as

$$P_s^*(\varepsilon) = \sqrt{\frac{-a_{3333} + \sqrt{a_{3333}^2 - 3a_{333333}(a_{33} - q_{1133}(\varepsilon_{11} + \varepsilon_{22}) - q_{3333}\varepsilon_{33})}}{3a_{333333}}} \quad (24)$$

Note that the higher order dielectric stiffness coefficients can be written with only one coefficient $a_{33}$ unknown for the entire composition range as shown in equations (10), (11) and (20). From the spontaneous polarization, the free energy barrier for the double well is expressed as

$$f_b = f(0) - f(P_s^*) \quad (25)$$

Note that the free energy barrier $f_b$ represents the polarization switching barrier at a fixed strain for the entire event. This value tends to be higher than that in actual stress-free switching due to the larger elastic energy contribution during switching. Nevertheless, the value should relate to the ferroelectric switching coercive field, and comparison across composition and strain states under identical conditions provides a trend.

The total strain $\varepsilon_{ij}$ is a sum of an elastic strain $\varepsilon^{ela}_{ij}$ and spontaneous strain $\varepsilon^0_{ij}$ under isothermal conditions (where there is no thermal strain contribution), namely,

$$\varepsilon_{ij} = \varepsilon_{ij}^{ela} + \varepsilon_{ij}^0 \quad (26)$$

Considering the relationship $\varepsilon^{ela}_{ij}$ to $P^*_s(\varepsilon)$ is more useful for intuitive understanding of the strain – spontaneous polarization relationship because the measured strain is an elastic strain added to already spontaneously strained states. From equations (3), (24) and (26), an elastic strain tensor component is expressed as,

$$\varepsilon_{ij}^{ela} = \varepsilon_{ij} - Q_{ij33} \frac{-a_{3333} + \sqrt{a_{3333}^2 - 3a_{333333}(a_{33} - q_{1133}(\varepsilon_{11} + \varepsilon_{22}) - q_{3333}\varepsilon_{33})}}{3a_{333333}} \quad (27)$$

As expressed in equations (24), (25) and (27), spontaneous polarization, free energy barrier and elastic strain are uniquely determined at an arbitrary total strain state with an assumed $a_{33}$ value. Recall the assumption that no shear strain components exist to preserve $P_1 = P_2 = 0$ in the model.

Spontaneous polarization, the free energy barrier, and elastic strain are all functions of $a_{33}$, but precise values of $a_{33}$ remain unknown, so further discussion requires sensitivity analysis. Fig. 3(a) and (b) show the spontaneous polarization and free energy barrier as a function of $a_{33}$ across the bounds mentioned above at various elastic strain states for Sc content $x = 0.2$. With zero elastic strain, the spontaneous polarization naturally remains the same as the stress-free spontaneous polarization, consistent with equations (7)-(9). Elastic strains of +/-1 % $\varepsilon^{ela}_{33}$ and +/- 0.3 % $\varepsilon^{ela}_{11} = \varepsilon^{ela}_{22}$ produce ~2 % change in spontaneous polarization. By contrast, spontaneous polarization changes by ~0.06 % across the entire range of possible $a_{33}$ values described in equation (13) as seen in Fig. 3(a). Unlike spontaneous polarization, a significant increase in the free energy barrier to polarization reversal is observed as the absolute value of $a_{33}$ increases (Fig. 3(b)), which is consistent with Fig. 1(a). Although there is a significant elastic strain effect on the free energy barrier, the $a_{33}$ value is the dominant factor to determine the barrier magnitude. These results suggest that the unknown $a_{33}$ value is not significant for spontaneous polarization evaluation, but the $a_{33}$ value is critical for free energy barrier analysis. Thus, spontaneous polarization is evaluated across a range of composition no matter what $a_{33}$ values are, and the free energy barrier is investigated only at $x = 0.2$, where the $a_{33}$ value was determined above.

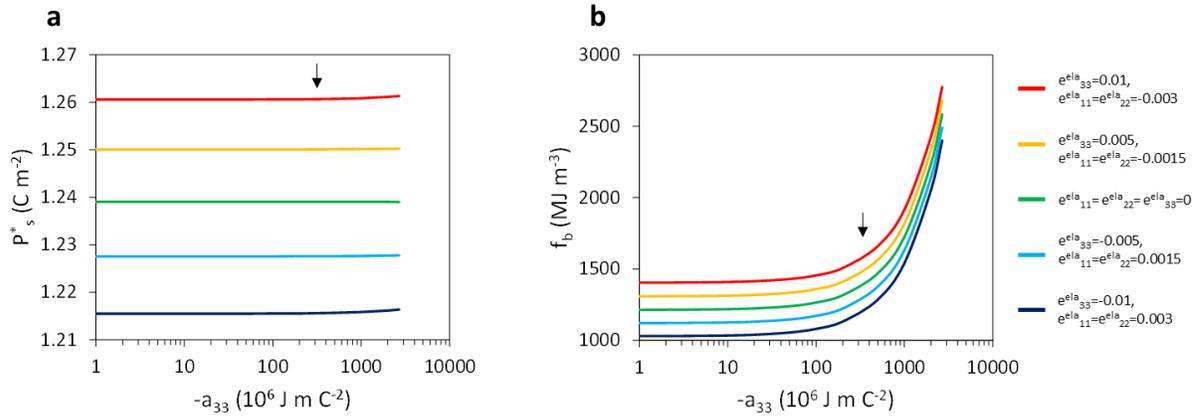

FIG. 3 Ferroelectric property sensitivity to values of the second order dielectric stiffness, $a_{33}$. (a) Spontaneous polarization and (b) free energy barrier as a function of $a_{33}$ under various elastic strain conditions. Black arrows indicate the $a_{33}$ value determined from published DFT calculations of barrier height.[23]

**B. Results and Discussion**

The existence of an elastic strain effect on spontaneous polarization is not surprising since strain – polarization coupling is common in ferroelectrics[6,43,44] and already reported in the wurtzite ferroelectrics.[14,16,26] However, our quantitative analysis provides important information about the magnitude of the effects of strain compared to other factors such as chemistry. Fig. 4(a) shows elastic strain – spontaneous polarization maps for $Al_{0.8}Sc_{0.2}N$ and $PbTiO_3$ for comparison. The coefficients of $PbTiO_3$ are from literature[45,46] after converting compliance from constant electric field $s^E$ to constant polarization $s^D$ for direct comparison to our model.[47] The spontaneous polarization values are normalized with the spontaneous polarization in a stress-free state. Both materials show larger spontaneous polarization at a normal strain with tensile 3 (polar) direction and compressive 1 and 2 directions, and smaller spontaneous polarization under a normal strain with compressive 3 (polar) direction and tensile 1 and 2 directions, which is intuitively reasonable. However, the degree of spontaneous polarization change under an elastic strain is different; $Al_{0.8}Sc_{0.2}N$ undergoes -2.7 to +2.5 % of change in spontaneous polarization under 1 % elastic strain, whereas $PbTiO_3$ exhibits -10 to +7.5 % of change in $P_s$ under the same percentage strain. This means that the polarization of $Al_{0.8}Sc_{0.2}N$ is less susceptible to elastic strain compared to $PbTiO_3$.

Fig. 4(b) shows the composition dependence of spontaneous polarization with envelopes representing maximum and minimum polarization values under ±1 % of $\varepsilon^{ela}_{33}$ and $\mp 0.3$ % of $\varepsilon^{ela}_{11} = \varepsilon^{ela}_{22}$ (see Fig. 3). The error bars along with the envelopes show the possible range due to the range of the $a_{33}$ value, which are quite small compared to the chemistry and strain effects discussed above. The strain corresponds to uniaxial strain along the 3 direction with assumed Poisson's ratio equal to 0.3. The plotted dots are the stress-free spontaneous polarization from DFT calculations.[32] From Sc content $x = 0.1$ to 0.4, a 14 % decrease in spontaneous polarization is seen, and this change is more than 5 times larger than the change attributed to ±1 % of uniaxial elastic strain along the polar direction. Note that 1 % of uniaxial elastic strain corresponds to 1.2 – 2.7 GPa, which is close to the fracture strength of AlN thin films[48] and more than the controllable range in residual stress with sputtering process variables.[49] Therefore, composition is more effective than lattice strain for controlling spontaneous polarization in $Al_{1-x}Sc_xN$. Lattice strain can tune spontaneous polarization up to a couple of percent, which is smaller than classic perovskite ferroelectrics $PbTiO_3$.

Fig. 4(c) illustrates the elastic strain – free energy barrier $f_b$ maps for $Al_{0.8}Sc_{0.2}N$ and $PbTiO_3$ for comparison. The $f_b$ values are normalized with the free energy barrier under the condition that the total strain is only spontaneous strain at a stress-free state. Both materials show a larger barrier height with larger spontaneous strain. In a range of ±1 % elastic strain, the elastic effect on $f_b$ is -19 to +20 % for $Al_{0.8}Sc_{0.2}N$, while that for $PbTiO_3$ is -42 to +45 %. Again, $Al_{0.8}Sc_{0.2}N$ shows smaller effects of elastic stain than does $PbTiO_3$, but the percent change in barrier height under an elastic strain is one order of magnitude larger than the percent change in spontaneous polarization, so the difference between the strain sensitivity between $Al_{1-x}Sc_xN$ and $PbTiO_3$ is significantly less for the barrier to polarization reversal than to the spontaneous polarization value itself.

Indeed, a recent experimental report suggested that the ionicity of Sc-N bonding is the dominant factor that controls spontaneous polarization in this system.[50] The result showing decreased spontaneous polarization with increasing Sc content even in an elastically strained film is consistent with Fig. 4(b). Coercive field, however, is commonly reported to be sensitive to elastic strain/stress,[14,16,26] consistent with the greater sensitivity of the free energy barrier to strain that we show here. A quantitative relationship between coercive field and thermodynamic calculations requires knowledge of the switching mechanism(s) and the intermediate state or static domain boundary configuration (e.g., the intermediate strain state during polarization switching attributed to a change in spontaneous strain as well as the elastic interaction across domain walls). Moreover, the intermediate state might possess in-plane polarization components i.e., $P_1$ and/or $P_2 \neq 0$, thereby requiring in-plane coupling and dielectric stiffness coefficients for further studies dealing with switching dynamics and domain evolution.

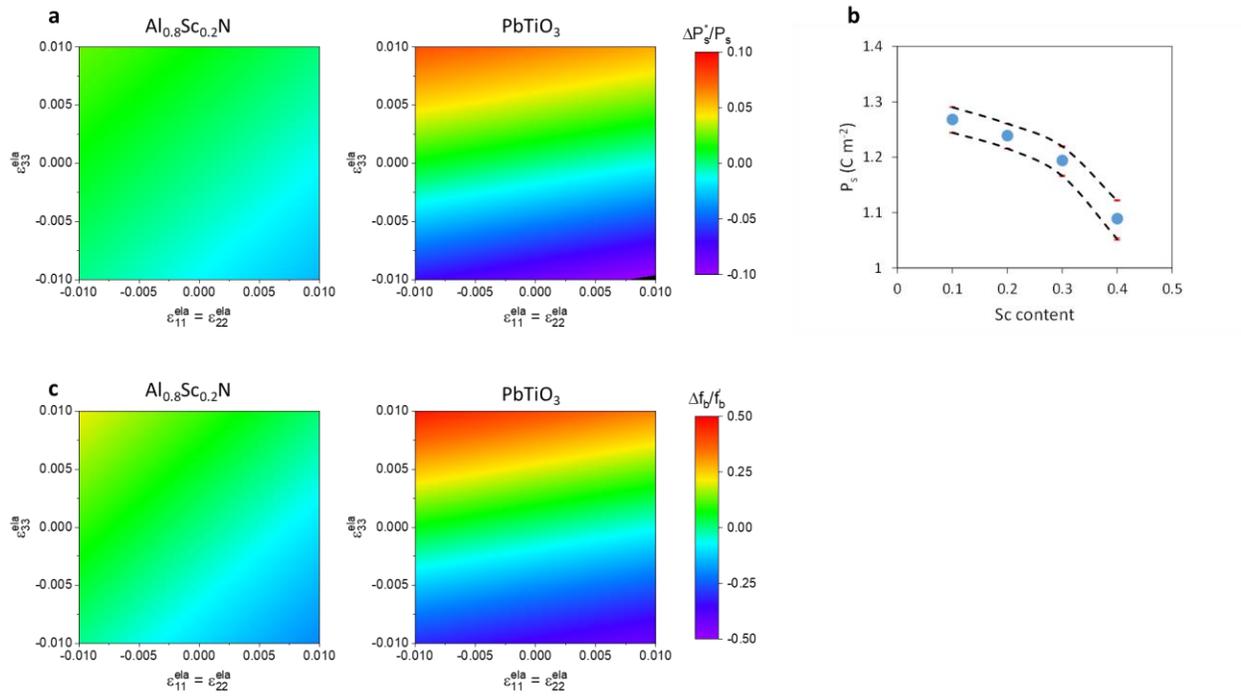

FIG. 4 Elastic strain effects on ferroelectric properties. (a) Normalized spontaneous polarization map for $Al_{0.8}Sc_{0.2}N$ and $PbTiO_3$ under ±1 % of elastic strain (b) Spontaneous polarization as a function of Sc content. Envelope (dashed lines) shows the maximum and minimum values under ±1 % of uniaxial elastic strain $\varepsilon^{ela}_{33}$. (c) Normalized free energy barrier map for $Al_{0.8}Sc_{0.2}N$ and $PbTiO_3$ under ±1 % of elastic strains.

## V. CONCLUSION

Free energy analysis of $Al_{1-x}Sc_xN$ with an electromechanical coupling term using the Landau – Devonshire formalism has been carried out based on reported DFT calculation results. Dielectric stiffness coefficients are determined based on the free energy curve shape under stress-free conditions, keeping the double well function for $x = 0 – 0.5$. A first order thermally-driven phase transition between the polar ferroelectric phase *6mm* and non-polar parent *6/mmm* phase is expected at $x = 0.2$ based on the dielectric stiffness coefficients. The electrostrictive coefficients for $x = 0 – 0.5$ are determined using reported piezoelectric coefficients, spontaneous polarization values, and dielectric susceptibility. The ambiguity due to the range of $a_{33}$ value is shown to be unimportant for the elastic strain – spontaneous polarization curve, and the elastic strain effect on spontaneous polarization is small compared to $PbTiO_3$. The elastic strain effect on free energy barrier height is also small for $Al_{1-x}Sc_xN$ compared to $PbTiO_3$, but the relative difference is not as large. These results explain the experimentally reported coercive field tunability under elastic strain/stress as well as the relative stability of spontaneous polarization with strain. Determination of the sensitivity of various properties to external variables such at strain provides crucial guidance for how to optimize properties for potential future applications.

**ACKOWLEDGEMENTS**

This work was co-authored by Colorado School of Mines and the National Renewable Energy Laboratory, operated by the Alliance for Sustainable Energy, LLC, for the U.S. Department of Energy (DOE) under Contract No. DE-AC36-08GO28308. Funding was provided by the DARPA Tunable Ferroelectric Nitrides (TUFEN) program (DARPA-PA-19-04-03. K.Y.),